\documentclass[pra,showpacs,twocolumn]{revtex4}
\usepackage{bm}
\usepackage{amsmath}
\usepackage{amssymb}
\usepackage{epsfig}

\newcommand{\eqb}{\begin{equation}}
\newcommand{\eqe}{\end{equation}}
\newcommand{\rd}{\mathrm{d}}
\newcommand{\bx}{{\bf x}}

\newcommand{\vare}{\varepsilon }
\newcommand{\pd}{\partial}

\begin{document}

\title{Fock-space quantum particle approach  for the two-mode boson model describing  BEC trapped in a double-well potential  }
\author{ V. S. Shchesnovich${}^1$ and M. Trippenbach${}^2$  }
\affiliation{ ${}^1$Universidade Federal do ABC, Santo Andr\'e  SP 09210-170
Brazil,
\\
$^2$Institute of Experimental Physics, Optics Division, Warsaw University, ul.
Hoz\.a 69, Warsaw 00-681, Poland }

\date{\today }

\begin{abstract}
We develop the Fock-space many-body quantum approach for large number of bosons,
when the bosons occupy significantly only few modes.  The approach is based on an
analogy with the dynamics of a single particle of either positive or negative mass
in a quantum potential, where the inverse number of bosons plays the role of Planck
constant. As application of the method we consider the Bose-Einstein condensate in
a double-well trap.  The ground state of positive mass particle corresponds to the
mean-field fixed point of lower energy, while that of negative mass to the excited
fixed point. In the case of attractive BEC above the threshold for symmetry
breaking, the ground state is a cat state and we relate this fact to the
double-well shape of the quantum potential for the positive mass quantum particle.
The tunneling energy splitting between the local Fock states of the cat state is
shown to be extremely small for the nonlinearity parameter just above the critical
value and exponentially decreasing with the number of atoms. In the repulsive case,
the phase locked ($\pi$-phase) macroscopic quantum self-trapping of BEC is related
to the double-well structure of the potential  for the negative mass quantum
particle. We also analyze the running-phase macroscopic quantum self-trapping state
and show that it is subject to quantum collapses and revivals. Moreover, the
quantum dynamics just before the first collapse of the running phase shows
exponential growth of dispersion of the quantum phase distribution which may
explain the growth of the phase fluctuations seen in the experiment on the
macroscopic quantum self-trapping.
\end{abstract}


\maketitle


\section{Introduction}

The fully analytical quantum treatment of a system of interacting $N$ bosons is a
hard problem, especially if the number of particles is large,  which is usually the
case of Bose-Einstein condensate (BEC). In the limit $N\gg 1$  the mean-field
approach leads to the Gross-Pitaevskii equation for the order parameter
\cite{BGL,PS,CD}. The mean-field limit may be termed ``the classical limit'' though
the Gross-Pitaevskii equation describes non-classical effects, for instance, the
quantum coherence of BEC, and reduces to the single-particle Schr\"odinger equation
in the limit of vanishing atomic interaction.

The relation between the full quantum and mean-field dynamics of BEC is a subject
of intensive studies.  For instance, manifestation of the classical bifurcation in
the quantum spectrum of a dimer was studied in Ref. \cite{AFKO}. The mean-field
dynamics of a two-mode model is shown to be modulated by the quantum collapses and
revivals \cite{MCWW}. In Ref. \cite{VA,AV} the Bogoliubov-Born-Green-Kirkwood-Yvon
hierarchy for statistical operators truncated at the second level was used and it
was pointed out that the many-body quantum corrections to the mean-field theory can
be identified as decoherence. The correspondence between the many-body quantum
dynamics and mean-field (we will also refer to it as a quantum classical
correspondence)  was also studied in the phase-space by employing Husimi
distributions \cite{QHAM}. The classical-quantum correspondence in the context of
quantum collapse and revivals in the nonlinear two and three mode boson models was
presented in \cite{SK2,SK3}, where authors described the BEC interband tunneling
between the high-symmetry points of an optical lattice. It was demonstrated that
the quantum evolution distinguishes between the stable and unstable classical fixed
points, following the classical dynamics about the stable points and diverging from
it at the unstable ones.  A three-mode boson model was treated in Ref. \cite{MJ}
where a visualization of the eigenstates of the quantum system referring to the
underlying classical dynamics was studied.

The purpose of this paper is to propose an  analytical method to study the
correspondence between the classical and quantum dynamics. We present the relation
between the classical fixed points and the eigenenergy states of the full quantum
Hamiltonian in the limit of large number of bosons. Our model is based on the
simple fact that for $N\gg1$ the $s$-mode boson model is equivalent to a quantum
mechanical dynamics of a single particle living in a compact $(s-1)$-dimensional
space. We point out that various quantum features of the system of large number of
bosons, not accessible within the mean-field limit, can be derived by the presented
method. For instance, one can describe quantum fluctuations, BEC fragmentation
(see, for instance, Ref. \cite{BECfrag}) and the cat-states formation in the
supercritical attractive case \cite{BECcats1,BECcats2}.

Our method resembles the classical WKB approach to discrete Schr\"odinger equation,
previously developed in another context (see, for instance, the review \cite{Braun}
and the references therein). The discrete WKB  quantization method was already
applied to the two-mode boson model \cite{GK}. For instance, it was shown that the
semiclassical approximation reproduces the energy eigenvalues even for relatively
small $N$.

Our approach is developed in the Fock space, where we expand the exact very
complicated Schr\"odinger equation for the state of $N$-boson system in the small
parameter $h=1/N$, the effective Planck constant,  reducing it to a much simpler
approximate equation resembling that of a single quantum particle in a potential.

In general our method is applicable to $s$-mode boson models. Here we concentrate
on the two-mode boson model, reducing it to a one-dimensional Schr\"odinger
equation for the effective quantum particle. We consider the important two-mode
 model describing  BEC in a double-well potential, i.e. the boson Josephson
junction model (see, for instance, the reviews \cite{Legg,GO}). BEC in a double
well trap  is a subject of fundamental experiments and has many potential
applications. For instance, it was used for direct observation of the macroscopic
quantum tunneling and the nonlinear self-trapping \cite{TunTrap}, considered as an
implementation of the atomic Mach-Zehnder interferometer \cite{AtInt,ThrBif}, as a
sensitive weak force detector \cite{JM,ChipSens} and for preparation of atomic
number squeezed states  for possible atomic interferometry on a chip \cite{AtSqz}.

For the two-mode model we find two different Schr\"odinger equations for
description of quantum dynamics about the classical fixed points, corresponding to
positive and negative mass of the effective quantum particle. Two reduced
Schr\"odinger equations appear due to the fact that a non-zero classical phase (of
the stationary point, in our case) leads to a singularity of the limit $h\to0$ and,
as the result, it appears in the reduced Schr\"odinger equation. Previously, this
has been accounted for by dividing the  turning points in the discrete WKB approach
into the usual and ``unusual'' (for details see Ref. \cite{Braun}).

The paper is organized as follows. In section \ref{secII} we  derive the effective
particle representation of the quantum boson Josephson junction model.   The two
resulting Schr\"odinger equations for the effective quantum particle are studied in
section \ref{secIII}. The classical stationary points and their stability
properties are studied from the quantum mechanics point of view in section
\ref{secIV}. In section \ref{secV} we derive the ground state of the quantum model.
In section \ref{secVI} the macroscopic quantum self-trapping (MQST) phenomenon is
related to the negative mass effective quantum particle. We also study the quantum
dynamics of the running phase MQST states. Section \ref{secVII} contains discussion
of the main results. Some of the auxiliary calculations are relegated to appendices
A, B, C and D.


\section{Quantum particle approach for  BEC in a double-well trap}
\label{secII}

Consider, for simplicity, the case of sufficiently weak atomic interaction, when
only the degenerate energy levels of the double-well trap  are occupied by BEC (we
consider the weakly asymmetric double-well with the first two energy levels being
quasi-degenerate, $\delta E = E_2-E_1\ll E_3-E_2=\Delta E$). Under this condition
one arrives at the reduced Schr\"odinger equation (see also appendix \ref{appH} and
Refs. \cite{AB,GO})
\begin{eqnarray}
&& \frac{i}{N}\pd_\tau|\psi\rangle = \hat{H}|\psi\rangle,\label{EQ1}\\
&& \hat{H}\equiv \left\{-\frac{1}{N}(a_1^\dag a_2  +
a_2^\dag a_1) +\frac{\vare}{N}n_1 +\frac{\gamma}{N^2}(n_1^2
+n_2^2)\right\},\nonumber
\end{eqnarray}
where $|\psi\rangle$ is the system state, $n_j = a^\dag_ja_j$ (the boson field
operator is expanded over the localized states $\Psi = \varphi_L(\bx)a_1
+\varphi_R(\bx)a_2$) and the time is measured in the tunneling time units: $\tau =
t/T$. The parameters of the model  are defined as
\begin{eqnarray}
&&T = \frac{2\hbar}{\delta E}, \quad \vare = \frac{2\delta V}{\delta E},\quad
\gamma = \frac{gN}{\delta E}\int\rd^3\bx\,\varphi_{L,R}^4,
\nonumber \\
&&\delta E =-2\int\rd^3\bx\,\varphi_L\left(-\frac{\hbar^2}{2m}\nabla^2
+V(\bx)\right)\varphi_R,\nonumber\\
&& \delta V = \int\rd^3\bx\,\left[\varphi_LV(\bx)\varphi_L - \varphi_RV(\bx)\varphi_R \right].
\label{EQ2}\end{eqnarray}
where $g=4\pi\hbar a_s/m$ with $a_s$ being the $s$-wave scattering length, $V(\bx)$
is the double-well potential (centered at zero, for below), the localized  states
$\varphi_L(\bx)$ and $\varphi_R(\bx)$  are given by the normalized sum and
difference of the ground state and the first excited state of the ``symmetrized''
double-well potential $V_s(\bx)\equiv [V(\bx) +V(-\bx)]/2$ (due to the symmetry the
integrals of the fourth power are equal; for more details, see Appendix in Ref.
\cite{soldw}).

The dimensionless parameter  $\delta V$ can be of the same order as $\delta E$,
thus $\vare$ can be arbitrary. Note that   $\gamma $ is finite (not small) and  the
applicability condition that the interaction energy is much less than the average
trap  energy spacing, i.e. $\gamma \ll \Delta E/\delta E$, is satisfied due to the
large r.h.s..

Subtracting the conserved quantity $\frac{\gamma}{2N^2}(n_1+n_2)^2$ from the
Hamiltonian in Eq. (\ref{EQ1}) we obtain the standard representation of the
Hamiltonian for the boson Josephson junction (see, for instance, Refs.
\cite{QHAM,GO}). Our interaction parameter is related to the standard parameters as
follows $\gamma = E_cN^2/E_j$.

We have divided the Schr\"odinger equation (\ref{EQ1}) by $N$ and defined the
parameters by extracting the explicit $N$-factors reflecting the ``order" of the
respective boson operators (e.g. $a^\dag_1a_1 \sim N$). This is used in the
effective quantum particle representation of Eq. (\ref{EQ1}) (see Eq. (\ref{EQ5})).
It is important to observe for the following (and the validity of the explicit
$N$-factors) that the physical parameters of the model are $N$-independent in the
thermodynamic limit, which is $N\to\infty$ at a constant density
$\frac{N}{\Omega}$, where $\Omega$ is the volume. Indeed,  $T$ and $\vare$ are
evidently constant, while the nonlinear parameter $\gamma \sim \frac{gN}{\delta
E\Omega}$ since $|\varphi_\alpha|^4 \sim \Omega^{-2}$.

On the other hand,  the factor $N^{-1}$ at the time derivative on the l.h.s. of Eq.
(\ref{EQ1}) plays the r\^ole of an effective Planck constant and the thermodynamic
limit $N\to \infty$ now acquires another meaning, namely it is the semiclassical
limit of Eq. (\ref{EQ1}). To make this clearer, let us rewrite Eq. (\ref{EQ1}) in
the explicit form using the Fock basis
\eqb
|\psi\rangle = \sum_{k=0}^NC_k|k,N-k\rangle,
\label{EQ3}\eqe
where the occupation number  corresponding to the left well is denoted by $k$. We
get
\eqb
\frac{i}{N}\frac{\rd}{\rd \tau}C_k = -\left(b_{k-1}C_{k-1} + b_{k}C_{k+1}\right)
+a_kC_k,
\label{EQ4}\eqe
with
\begin{eqnarray*}
&& b_k = N^{-1}\left[(k+1)(N-k)\right]^{\frac12},\nonumber\\
&& a_k =  \gamma N^{-2}\left[k^2 + (N-k)^2\right]+\vare N^{-1}k.
\end{eqnarray*}
Introducing a continuous variable $x = k/N$, the effective Planck constant $h =
1/N$ and a wave-function $\Psi(x) = \sqrt{N}C_k$ we obtain Eq. (\ref{EQ1}) in yet
another form
\eqb
ih\pd_\tau\Psi(x) = -\left\{e^{-i\hat{p}}b_h(x) + b_h(x)e^{i\hat{p}}\right\}\Psi(x)
+ a(x)\Psi(x),
\label{EQ5}\eqe
with  $\hat{p} = -ih\pd_x$ and
\[
b_h(x) = [(x+h)(1-x)]^{1/2},\quad a(x) = \gamma[x^2 +(1-x)^2]+\vare x.
\]
Here $x\in(0,1)$ and $\hat{p}$ are  canonical variables with the usual commutator
$[\hat{p},x] = -ih$ \footnote{We suppose that the wave-function $\Psi(x)$ is zero
at the boundaries of the interval $[0,1]$, as is in all the cases in this paper.
The difficulties with defining the conjugated phase and amplitude variables in the
quantum case  are well-known \cite{PhaseAmpl}.}. Note that the inner product for
$N\gg1$ reads \mbox{$\langle\Psi_1|\Psi_2\rangle = \int_{0}^1\rd
x\,\Psi^*_1(x)\Psi_2(x)$}.

Eq. (\ref{EQ5}) is written in the $x$-representation form. It can be put in the
$\phi$-representation by putting $\hat{p} \to \phi$ and $x-1/2 \to ih\partial_\phi$
(the shifted variable is more convenient). In a more rigorous way, one can use the
transformation of  the state vector  $C_k$ to the Fourier space, given in our case
by the discrete Fourier transform,
\eqb
C_k = \frac{1}{\sqrt{N+1}}\sum_{l=0}^Ne^{-ik\phi_l}\tilde{C}_l,\quad \tilde{C}_l
=\frac{1}{\sqrt{N+1}}\sum_{k=0}^N e^{ik\phi_l}C_k,
\label{DFT}\eqe
where $\phi_l = 2\pi l/(N+1)$. Using periodicity of the exponent we have for $N
\gg1 $
\eqb
\Psi(x) = \frac{1}{2\pi}\int\limits_{-\pi}^{\pi}\rd \phi\, e^{iN\phi
x}\tilde{\Psi}(\phi),\; \tilde{\Psi}(\phi) = N\int\limits_{0}^{1}\rd x\, e^{-iN\phi
x}\Psi(x),
\label{FTR}\eqe
where $\tilde{\Psi}(\phi_l) = (N+1)\tilde{C}_l$. Thus, for $N\gg1$ the Hamiltonian
for boson Josephson junction in the $\phi$-representation reads (save an
inessential constant term)
\begin{eqnarray}
\hat{H} &=& -2\gamma h^2\pd_\phi^2  - \{b_h(1/2+ih\pd_\phi),\cos(\phi)\}
\nonumber\\
& & +  i\vare h \pd_\phi - i[b_h(1/2+ih\pd_\phi),\sin\phi]
\label{H4BJJ}\end{eqnarray}
where $\{ ...\}$ denote the anti-commutator. The Hamiltonian (\ref{H4BJJ}) reduces
to the usual non-rigid pendulum  form \cite{MFDW} if  one  neglects the variation
of $b_h(x)$ with respect to $x$.

 The limit of  large $N$ of Eq. (\ref{EQ5}) can be considered  by the WKB approach
for the discrete Schr\"odinger equation \cite{Braun}. We write the solution of Eq.
(\ref{EQ5}) in the form $\Psi = \exp\{iS(x,\tau,h)/h\}$ where $S(x,\tau,h)$ is
understood as a series $S = S^{(0)}(x,\tau) + hS^{(1)}(x,\tau) + O(h^2)$ and  the
terms are assumed to be differentiable functions. In the lowest order we obtain
from Eq. (\ref{EQ5}) the Hamilton-Jacobi equation for the classical action
$S^{(0)}(x,\tau)$:
\eqb
-S^{(0)}_\tau =  a(x) - 2b_0(x)\cos\left(\frac{\pd S^{(0)}}{\pd x}\right),
\label{EQ6}\eqe
where $b_0(x) = \sqrt{x(1-x)}$. The classical Hamiltonian obtained from Eq.
(\ref{EQ6}) reads
\eqb
\mathcal{H} = a(x) - 2b_0(x)\cos\phi
\label{CHAM}\eqe
with the canonical variables $x$ and $\phi\in(0,2\pi)$: $\{\phi,x\}=1$ (cf. with
Ref. \cite{MFDW}, where $z = 1-2x$). The first-order approximation reads $S^{(1)} =
\frac{i}{2}\ln [b_0(x)\sin(S^{(0)}_x)]$.

Let us, however, take  a slightly different approach. The most important features
of the classical dynamics are the stationary points and their stability properties.
About a stationary point (say, $x_s$ and $\phi_s$) the classical action can be
expanded as follows
\eqb
S^{(0)}(x,\tau) = -E^{(cl)}\tau + \phi_s(x - x_s) + O[(x-x_s)^2],
\label{EQ7}\eqe
where we have used that $S^{(0)}(x,\tau) = -E^{(cl)}\tau +S^{(st)}(x)$ and
$\frac{\pd S^{(st)}}{\pd x}(x_s) = \phi_s$. We are interested in the limit of large
$N$. The wave-function localized about some $x_s$-point has a singular derivative
in the limit $h\to 0$: $\frac{\pd \Psi}{\pd x} \sim ih^{-1}\phi_s\Psi(x,\tau) $ if
$\phi_s\ne0$. The singularity is due to the classical phase in the wave-function.
We then account for the phase explicitly, introducing the transformation (at some
$x_s$-point)
\eqb
\Psi(x,\tau) = e^{iN\phi(\tau) (x-x_s)}\psi(x,\tau)
\label{EQ8}\eqe
before taking the limit of large $N$.  The phase $\phi(\tau)$ satisfies the
evolution equation $\dot{\phi} = \pd H(x_s,\phi)/ \pd x$ (by expanding the
Schr\"odinger equation in $x-x_s$ and taking the limit $h\to0$), hence it is
nothing but the classical phase.

To obtain a reduced Schr\"odinger equation we substitute the representation
(\ref{EQ8}) into Eq. (\ref{EQ5}) and expand the result into series in $\hat{p}$
keeping the terms up to $O(\hat{p}^2)$ (since the derivative is regular the
expansion is actually in $h$). We have
\begin{widetext}
\begin{eqnarray}
\left\{e^{-i\hat{p}}b_h(x) +b_h(x)e^{i\hat{p}}\right\}e^{i\frac{\phi}{h}x}\psi(x)
&=& e^{i\frac{\phi}{h}x}\biggl\{2\cos(\phi) b_h(x) - h\cos(\phi)\frac{\rd b_h}{\rd
x} - \sin\phi\left[\hat{p}b_h(x) + b_h(x)\hat{p} \right]\nonumber\\
&&-\frac12\left[e^{-i\phi}\hat{p}^2b_h(x)
+e^{i\phi}b_h(x)\hat{p}^2\right]+O(\hat{p}^3)\biggr\}\psi(x).
\label{EQ9}\end{eqnarray}
\end{widetext}
One can make one more reduction  in Eq. (\ref{EQ9}) by using that
\begin{eqnarray}
&&\frac12\left[e^{-i\phi}\hat{p}^2b_h +e^{i\phi}b_h\hat{p}^2\right]= \cos(\phi)\hat{p}b_h\hat{p}
\nonumber\\
&&- \frac12\left\{h\sin\phi\left(\frac{\rd b_h}{\rd x}\hat{p}+\hat{p}\frac{\rd
b_h}{\rd x}\right) + \frac{h^2}{2}\cos(\phi)\frac{\rd^2 b_h}{\rd x^2}
\right\}.\qquad
\label{EQ10}\end{eqnarray}
Using Eqs. (\ref{EQ9}) and (\ref{EQ10}) in  the Schr\"odinger equation (\ref{EQ5})
we obtain
\begin{widetext}
\eqb
ih\pd_\tau\psi = \mathcal{V}_h(x,\phi)\psi  + \Bigl[\cos\phi\,\hat{p}b_h(x)\hat{p}
+ \sin\phi\left(b_h(x)\hat{p}+\hat{p}b_h(x)\right)
-\frac{h}{2}\sin\phi\left(\frac{\rd b_h}{\rd x}\hat{p}+\hat{p}\frac{\rd b_h}{\rd
x}\right)\psi  +O(\hat{p}^3)\Bigr]\psi,
\label{EQ12}\eqe
\end{widetext}
where the potential $V_h$ reads
\eqb
\mathcal{V}_h =  a(x) - 2b_h(x)\cos\phi + h\cos\phi\left[\frac{\rd b_h}{\rd x}
-\frac{h}{4}\frac{\rd^2 b_h}{\rd x^2}\right].
\label{Vh}\eqe
Reducing further, one can neglect the terms of order $O(h)$ in the potential
$\mathcal{V}_h$, since it contains the $h$-independent term, obtaining nothing but
 the classical Hamiltonian (\ref{CHAM}) $\mathcal{V}_0(x) =\mathcal{H} = a(x)
-2b_0(x)\cos\phi$.

The analogy with the quantum problem of a particle in a  potential becomes evident
from the Schr\"odinger equation in the form of Eq. (\ref{EQ12}).  The stationary
states $\psi(x,\tau) = e^{-iE\tau/h}\psi_E(x)$ correspond to the extrema of the
quantum potential. The extremal values of the classical phase (which is a
parameter) given by $\frac{\pd \mathcal{V}_0(x,\phi)}{\pd \phi}=0$ coincide with
the classical values at the stationary points: $\phi = 0$ or $\phi = \pi$. These
correspond to positive or negative mass of the effective quantum particle.


\section{Quantum model about the mean-field stationary points}
\label{secIII}

The  Schr\"odinger equation (\ref{EQ12}) in the two cases $\phi_+=0$ and $\phi_- =
\pi$  reduces to
\eqb
ih\pd_\tau\psi =  \left\{ \pm\hat{p}\sqrt{x(1-x)}\hat{p}
+\mathcal{V}_\pm(x)\right\} \psi,
\label{EQ14}\eqe
where the upper/lower sign corresponds to $\phi_+/\phi_-$ and $\mathcal{V}_\pm(x) =
\gamma [x^2 +(1-x)^2] +\vare x \mp2\sqrt{x(1-x)}$. For further analysis it is
convenient to introduce a new variable $z = 1 - 2x$, the  atomic population
difference   divided by the total number of atoms. We get
\eqb
ih\pd_\tau\psi =  \left\{ \mp 2h^2\frac{\rd }{\rd z}\sqrt{1-z^2}\frac{\rd }{\rd z}
+V_\pm(z)\right\} \psi,
\label{EQ15}\eqe
where the potential simplifies to
\eqb
V_\pm (z) = \frac12(\gamma{}z^2 - \vare{}z) \mp\sqrt{1-z^2}.
\label{EQ16}\eqe
Note that the potential $V_\pm(z)$  crucially depends on the nonlinearity parameter
$\gamma$.

Taking into account the symmetry of Eq. (\ref{EQ15}) with respect to replacing
$\pm$ we have the following mapping between the repulsive and attractive BEC cases:
\begin{eqnarray}
&&\gamma_\mathrm{attr} = -\gamma_\mathrm{rep},\quad z_\mathrm{attr} =
-z_\mathrm{rep}, \nonumber\\
&& \psi_\mathrm{attr}(z,\tau;\gamma,\phi_\pm) =
\psi^*_\mathrm{rep}(-z,\tau;-\gamma,\phi_\mp),
\label{RepAttr}\end{eqnarray}
i.e., for instance,  an attractive BEC with positive mass $\phi=0$ is equivalent to
a repulsive BEC with negative mass $\phi=\pi$ evolving backwards in time in the
inverted space $z\to-z$. Note that it is sufficient to consider also $\vare\ge0$,
since $\vare<0$ can be simulated by replacing $z\to-z$.

\begin{figure}[htb]
\begin{center}
\psfig{file=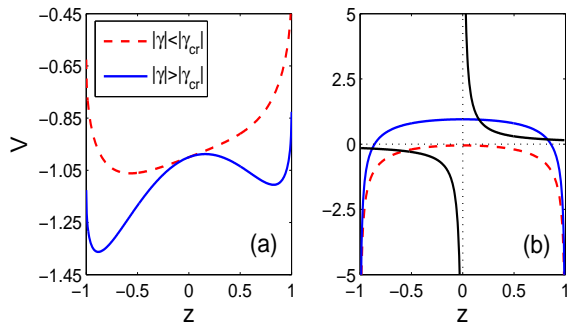,height=0.25\textwidth,width=0.425\textwidth} \caption{(Color
online) The potential for the effective quantum particle, panel (a), (here $V
=-V_-(z)$ for the repulsive BEC corresponding to negative mass, $\phi=\pi$, or $V
=V_+(-z)$ for the attractive BEC for $\phi=0$, positive mass) and the graphical
solution of Eq. (\ref{EQ17}), panel (b) (again for  $\phi=\pi$ for the repulsive
BEC, or for $\phi=0$ for the attractive BEC, in the last case the axes are
inverted). Here $\vare = 0.3$ which gives $|\gamma_{\mathrm{cr}}| = 1.4521$. }
\label{FG1}
\end{center}
\end{figure}

The  equivalence (\ref{RepAttr}), however, is useful only in the general analytical
analysis, the attractive BEC is fundamentally different from the repulsive BEC. As
we show below, in both cases the Schr\"odinger equation (\ref{EQ15}) with the
positive mass gives the ground state of BEC and only in the attractive case
$V_+(z)$ has the double-well form (see Fig. \ref{FG1}(a) below).

The  extrema of the potential (\ref{EQ16}) solve the equation
\eqb
\gamma \pm \frac{1}{\sqrt{1-z^2}} = \frac{\vare}{2z}.
\label{EQ17}\eqe
Consider $\gamma>0$, $\vare\ge 0$. In the case of positive mass ($\phi=0$) there is
just one minimum. For the case of negative mass ($\phi=\pi$) there is a critical
$\gamma_{\mathrm{cr}}$ below which there is just one solution which is a minimum
of $V=-V_-(z)$, whereas for $\gamma>\gamma_{\mathrm{cr}}$ there are three
solutions, two corresponding to local minima and one to a local maximum, see Fig.
\ref{FG1}(a). Eq. (\ref{EQ17}) is easily solved graphically by plotting the l.h.s.
and the r.h.s. as functions of $z$, see Fig. \ref{FG1}(b). Hence, in the case of
$\phi=0$ (the upper sign) we get that the positive-mass effective quantum particle
moves in a single well potential $V=V_+(z)$ and there is only one solution to Eq.
(\ref{EQ17}). In the case of $\phi = \pi$ the negative-mass particle sees either
one well or two well potential $V=-V_-(z)$.

The critical value of the nonlinearity coefficient can be found by equating the
derivatives of the two sides of Eq. (\ref{EQ17}), which gives an additional
equation. We get
\eqb \gamma_{\mathrm{cr}} =
\left[1+\left(\frac{\vare}{2}\right)^{\frac23}\right]^{\frac32},
\label{EQ18}\eqe
while the corresponding solution, i.e. the point of inflection, reads
\mbox{$z_{\mathrm{cr}} = [1+(2/\vare)^{2/3}]^{-1/2}$}.  For
$\gamma>\gamma_{\mathrm{cr}}$  two positive solutions bifurcate from it:
$z_1<z_{\mathrm{cr}}<z_2$, where $z_1$ is the local maximum of the inverted
potential $-V_-(z)$. The local maximum  solution is an analog of the $z=0$ solution
for $\vare \ne 0$.

One can show (see  Appendix \ref{appA}) that in the case of $\phi=\pi$ the negative
local minimum of the inverted quantum potential is also the absolute minimum (of
this potential, i.e. $-V_-(z)$) for all $\gamma>\gamma_\mathrm{cr}$ and
$\vare\ge0$, whereas the negative $z$ correspond to higher atomic population of the
trap well with the higher zero-point energy (see Eq. (\ref{EQ1})).  This fact,
however, is easily explained  within our approach  by the negative mass of the
effective quantum particle.

Finally, all the results obtained for the repulsive BEC are transferred to the
attractive BEC case by using the equivalence given in Eq. (\ref{RepAttr}). For
instance the cases of $\phi_\pm$ are interchanged and  the double-well potential of
Fig. \ref{FG1}(a) in the case of attractive BEC appears for $\phi=0$ (i.e. now $V =
V_+(-z)$) and \mbox{$\gamma<-\gamma_\mathrm{cr}
=-\left[1+\left(\frac{\vare}{2}\right)^{2/3}\right]^{3/2}$.}


\section{The mean-field stationary points and their stability explained  }
\label{secIV}

\begin{figure}[htb]
\begin{center}
\psfig{file=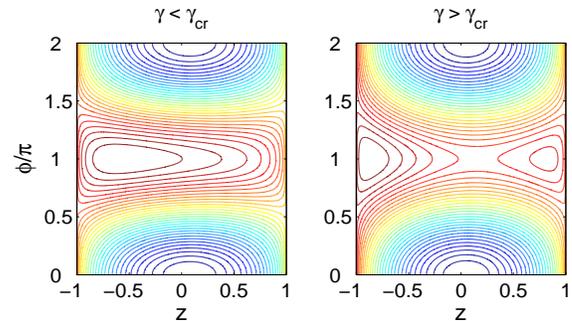,height=0.25\textwidth,width=0.425\textwidth} \caption{(Color
online) The phase portrait of the classical Hamiltonian Eq. (\ref{EQ19})
corresponding to a repulsive BEC in a double-well trap. The parameters are as in
Fig. \ref{FG1}. }
\label{FG2}
\end{center}
\end{figure}

Before we proceed with the analysis of the bound state wave-functions, let us
explain from the quantum mechanical point of view  the appearance and stability
properties of the stationary points of the classical Hamiltonian (see also Ref.
\cite{MFDW}). In the new variables we have
\eqb
\mathcal{H} = \frac{\gamma}{2}z^2 -\frac{\vare}{2}z-\sqrt{1-z^2}\cos\phi,
\label{EQ19}\eqe
where one must remember that $\{\phi,z\}= -2$. The stationary points of the
classical Hamilton equations, $\dot{z}=-2\frac{\pd \mathcal{H}}{\pd \phi}$,
$\dot{\phi} = 2\frac{\pd \mathcal{H}}{\pd z}$, correspond to  the extrema of the
quantum potential ${V}(z)$, i.e. $\cos\phi_s=\pm1$, whereas $z_s$ is defined from
Eq. (\ref{EQ17}). Consider the case $\gamma>0$ (repulsive BEC) and $\vare \ge 0$.
The stability properties are defined by the local Hamiltonian, see also Fig.
\ref{FG2}. Introduce the local phase space co-ordinates $\phi = \phi_s+\varphi$ and
$z = z_s +\zeta$.

1. In the case of $\phi_s=0$ we have up to the second-order terms in the local
variables
\eqb
\mathcal{H}_+ = V_+(z_s) + \sqrt{1-z_s^2}\frac{\varphi^2}{2}+\left[\gamma
+(1-z_s^2)^{-3/2}\right]\frac{\zeta^2}{2},
\label{EQ20}\eqe
hence the classical stationary point $\phi_s=0$ and $z=z_s$, solving Eq.
(\ref{EQ17}) for the upper sign, is elliptic and stable. This stationary point, in
fact, corresponds to the absolute minimum of the quantum potential $V_+(z)$.

2. In the case of $\phi_s=\pi$ we obtain up to the second-order terms (by simply
changing the sign at $\cos\phi$)
\eqb
\mathcal{H}_- = V_-(z_s)
-\sqrt{1-z_s^2}\frac{\varphi^2}{2}-\left[(1-z_s^2)^{-3/2}-\gamma
\right]\frac{\zeta^2}{2}.
\label{EQ21}\eqe
First, consider the solution to Eq. (\ref{EQ17}) such that $z_s<0$.  Using Eq.
(\ref{EQ17}) in the following estimate
\begin{eqnarray*}
&&\gamma -\frac{1}{(1-z_s^2)^{3/2}}<\frac{\gamma}{1-z_s^2}
-\frac{1}{(1-z_s^2)^{3/2}} \\
&&=\frac{\vare}{2z_s(1-z_s^2)}<0,
\end{eqnarray*}
we conclude that the negative stationary point $z_s<0$ for $\phi_s = \pi$ is always
elliptic  and hence stable. It actually corresponds to the absolute minimum of
$-V_-(z)$ (see Appendix \ref{appA}).

Now consider the positive stationary points $z^{(1)}_s<z_\mathrm{cr}<z^{(2)}_s$,
which appear for $\gamma> \gamma_{\mathrm{cr}}$. For $z^{(1)}_s$ we have
\[
\gamma -\frac{1}{\left(1-[z^{(1)}_s]^2\right)^{3/2}}>\gamma
-\frac{1}{(1-z_\mathrm{cr}^2)^{3/2}}=\gamma-\gamma_\mathrm{cr}>0,
\]
where we have used that $\gamma_\mathrm{cr}= (1-z_\mathrm{cr}^2)^{-3/2}$. Thus this
stationary point is hyperbolic and hence unstable (recall that it corresponds to
the local maximum of $-V_-(z)$). The other stationary point
$z^{(2)}_s>z_\mathrm{cr}$ is a local minimum of the quantum potential $-V_-(z)$.
Let us show that it is always elliptic, i.e. stable. Indeed, using Eq. (\ref{EQ17})
we obtain the estimate ($z=z^{(2)}_s$)
\begin{eqnarray}
&&\gamma - \frac{1}{(1-z^2)^{3/2}} = \left(\gamma -
\frac{1}{\sqrt{1-z^2}}\right)\frac{1}{1-z^2} - \frac{\gamma z^2}{1-z^2}\nonumber\\
&& = \frac{1}{1-z^2}\left(\frac{\vare}{2z}-\gamma
z^2\right)=-\frac{z^2}{1-z^2}\left[\gamma -\frac{\vare}{2z^3}\right]<0,
\end{eqnarray}
where we have used that $z^{(2)}_s >z_\mathrm{cr}$, the expression
$\gamma_\mathrm{cr} = {\vare}/(2z^3_\mathrm{cr})$ and that
$\gamma>\gamma_\mathrm{cr}$.

According to Eq. (\ref{RepAttr}), the classical action for Eq. (\ref{EQ15})
satisfies $S^{(0)}_\mathrm{attr}(z,\tau) =- S^{(0)}_\mathrm{rep}(-z,\tau)$ where
$\gamma\to-\gamma$ and $\phi_\pm\to\phi_\mp$, thus  Eqs. (\ref{EQ20}) and
(\ref{EQ21}) can be transferred to an attractive BEC by the substitution:
$-\mathcal{H}_\pm\to \mathcal{H}_\mp$ and $-V_\pm \to V_\mp$.

In conclusion of this section, the local minima of the quantum potential (or the
inverted potential, in the case of negative mass) correspond to the elliptic
stationary points of the classical dynamics and the local maximum to a hyperbolic
stationary point, which is a physically clear result.


\section{The ground state of  BEC in a double-well trap}
\label{secV}

To find the ground state of BEC in the double-well trap one has  to analyze the
energy of the bound states localized at the extrema of the potentials $V_\pm(z)$.
For $N\gg1$ one needs to compare just the zero-point energies at the extremal
points (which is the classical energy of the stationary point, Eqs. (\ref{EQ20})
and (\ref{EQ21})). The result is that for both attractive and repulsive BEC in a
double-well trap the ground state is given by that of the effective quantum
particle with the positive mass, i.e. with the classical phase $\phi=0$ (see
appendix \ref{appB} for details).

The direct link between the  stable classical stationary points and the nature of
the  corresponding quantum states discussed in section \ref{secIV} allows one to
get the local approximation  for the quantum bound states by directly quantizing
the local Hamiltonian, Eqs. (\ref{EQ20}) and (\ref{EQ21}), by replacing $\varphi
\to 2ih\pd_\zeta$, since $[\hat{\varphi},\zeta] = 2ih$ (remembering that in the
case of attractive BEC Eq. (\ref{EQ20}) corresponds to $\phi=\pi$, while Eq.
(\ref{EQ21}) to $\phi=0$). We slightly correct this scheme in the negative mass
case, where the wave function $\Psi(x,t)$ of BEC in a double-well trap has a
non-trivial phase according to  Eq. (\ref{EQ8}).

Writing the local classical Hamiltonian in Eqs. (\ref{EQ20})-(\ref{EQ21}) as
\eqb
H_\pm(\zeta,\varphi) =V_\pm(z_s)+ (a_\pm\varphi^2 +b_\pm \zeta^2)/2,
\label{Hamab}\eqe
we obtain  an approximation  to the bound state of BEC corresponding to a stable
stationary point of the classical dynamics and its energy as follows:
\begin{eqnarray}
&& \Psi_{\mathcal{E}_\pm}(x) =
\frac{1}{\pi^{1/4}\sigma_\pm^{1/2}}\exp\left\{i\frac{\phi_\pm}{h}x
-\frac{(x - x_s)^2}{2\sigma_\pm^2}\right\},\nonumber\\
&&\mathcal{E}_\pm = V_\pm(z_s)+ O(h),
\label{EQ28}\end{eqnarray}
where $\sigma_\pm^2 = \frac{1}{2N}\sqrt{\frac{a_\pm}{b_\pm}}$ is the Gaussian width
and we have used that $\Delta z = 2\Delta x$. The  energy spacing for the few lower
levels reads $\Delta\mathcal{E}_\pm= h\Omega_\pm$, where $\Omega_\pm = 2\sqrt{a_\pm
b_\pm}$ is the classical frequency. In the case of negative mass the energy levels
of a local quantum Hamiltonian are descending.

\subsection{The ground state of repulsive BEC}
\label{secVA}

For $\gamma>0$  we obtain the single solution of Eq. (\ref{EQ17}) corresponding to
the positive mass  case in the form of a series in $\vare$:
\eqb
z_s = \frac{\vare}{2}(1+\gamma)^{-1}\left[ 1 -\frac{\vare^2}{8(1+\gamma)^3}
+O(\vare^4)\right],
\label{zs}\eqe
which is valid for a weakly asymmetric  potential $V_+(z)$, i.e.
$\vare\ll1+\gamma$. The average atomic population difference between the two wells
reads \mbox{$\langle n_1-n_2\rangle = z_sN$.} By the substitution
$\gamma\to-\gamma$ and $z_s\to-z_s$ Eq. (\ref{zs}) also gives the position of the
minimum of the inverted potential $-V_-(z)$ in the subcritical case
$\gamma<\gamma_\mathrm{cr}$.

For a weakly asymmetric  potential $V_+(z)$ one can also calculate the atom number
fluctuations  using the local approximation for the ground state given by Eq.
(\ref{EQ28}) ($x_s = (1-z_s)/2$):
\begin{eqnarray}
&& \langle\Delta n_1^2\rangle = N^2\langle(x-x_s)^2\rangle = \frac{N^2\sigma_+^2}{2}   \nonumber\\
&&\quad =
\frac{N}{4}(1+\gamma)^{-1/2}\left[1-\frac{\vare^2(4+\gamma)}{16(1+\gamma)^3}
+O(\vare^4)\right],\quad
\label{EQ29}\end{eqnarray}
where we have used that
\eqb
\sigma^2_+ = \frac{1}{2N}\left[\frac{(1-z_s)^{1/2}}{\gamma +
(1-z_s^2)^{-3/2}}\right]^{1/2}
\label{sigma+}\eqe
and the series expression (\ref{zs}) (see the details in Appendix \ref{appC}).

The result given by Eqs. (\ref{EQ28}) and (\ref{EQ29}) can be compared to the
non-interacting case, which is exactly solvable (here we consider the case $\vare =
0$). Indeed,  the tunneling term of the quantum Hamiltonian (\ref{EQ1}) can be
diagonalized by the canonical transformation $c_\pm = (a_1 \pm a_2)/\sqrt{2}$ with
the effect $H = -(a_1^\dag a_2^{} +a_2^\dag a_1^{})/N = ( c_-^\dag c_-^{}-c_+^\dag
c_+^{})/N$. The eigenstates are given by $(c_+^\dag)^m(c_-^\dag)^{N-k}|0,0\rangle$,
where $a_1|0,0\rangle =  a_2|0,0\rangle = 0$. The ground state corresponds to $m=N$
\eqb
|\Phi\rangle = \frac{1}{\sqrt{2^N N!}}\left(a_1^\dag  + a_2^\dag
\right)^N|0,0\rangle,
\label{EQ30}\eqe
For large $N$,  approximating the  factorial, we get  the coherent state in the
Fock basis as
\begin{eqnarray}
\Phi(z) &=& \sqrt{N}\langle\frac{1-z}{2}N,\frac{1+z}{2}N|\Phi\rangle\nonumber\\
&=& \left(\frac{2N}{\pi}\right)^{\frac{1}{4}}\exp\left\{-\frac{N}{4}\left(z^2 +
\frac{z^4}{6}+...\right)\right\}.
\label{EQ31}\end{eqnarray}

Eqs. (\ref{EQ28}) - (\ref{EQ29}) describe all three known regimes of repulsive BEC
tunneling in the double-well trap (see for instance, Ref. \cite{Legg,GO}): (1) Rabi
regime, when the coherence is very high and the atom number fluctuations are large
(essentially the interaction free regime), $\gamma\ll1$; (2) Josephson regime, when
the coherence is high and the atom number fluctuations are small, $1\ll \gamma \ll
N^2$ and (3) Fock regime, when the coherence is low and the atom number
fluctuations vanish, $\gamma \gg N^2$.

\subsection{The ground state of attractive BEC }
\label{secVB}

\textit{Subcritical $\gamma$.} In the case $|\gamma| < \gamma_{\mathrm{cr}}$ the
ground state of attractive BEC is essentially the same as that of the repulsive BEC
and is given by the general result (\ref{EQ28}) for the positive mass case (i.e.
the upper sign). The average population difference is again given by the series
solution (\ref{zs}) and the number fluctuations are given by the same formula
(\ref{EQ29}) as in the repulsive case. However, since now $\gamma<0$, there is an
essential difference: even moderate attractive interactions strongly enhance the
atom number fluctuations (which fact leads to absence  of the Josephson regime for
attractive BEC).

For $\gamma$ approaching (from above) the critical value $-\gamma_\mathrm{cr}$  the
fluctuations given by the local oscillator approximation  (\ref{EQ28}) diverge,
which is easily seen from Eq. (\ref{EQ29}) applied to the case $\vare=0$. In the
general case the result is similar and follows from the exact expression for
$\sigma^2_+$ with the use of the exact value for the point of inflection $z_s$
(given below Eq. (\ref{EQ18}) of section \ref{secIII}) and the expression
$\gamma_\mathrm{cr} = (1-z_\mathrm{cr}^2)^{-3/2}$ \footnote{ Note that Eq.
(\ref{EQ29}) seems to give a higher value $\gamma=-1$ than $-\gamma_\mathrm{cr}$
for divergence of the fluctuations for $\vare\ne0$. However, by more careful
inspection one notices that the second term in the square brackets for the number
fluctuations becomes of order one at the critical $\gamma$, this is an artefact  of
the approximation in Eq. (\ref{zs}).}. The divergence results from the oscillator
approximation of the wave-function $\Psi(x)$ (\ref{EQ28}) which breaks down at the
critical value of $\gamma$, since the potential $V_+(z)$  for
$\gamma=-\gamma_\mathrm{cr}$ is approximated about the stationary point $z=z_s$ by
the fourth order power in $z-z_s$ (we give the case $\vare=0$ and $z_s=0$  for
simplicity):
\eqb
V_+(z) = -\frac{ z^2}{2} - \sqrt{1-z^2} = \frac{z^4}{8} + \frac{z^6}{16} + ....
\eqe
Hence, the usual oscillator approximation should be replaced about the critical
$\gamma$ by the ground state in the non-harmonic potential, not known in  the exact
form.

\textit{Supercritical $\gamma$.} In the general case of
$|\gamma|>\gamma_\mathrm{cr}$ one has to solve the stationary Schr\"odinger
equation (\ref{EQ15}) to obtain the ground state. However, for $\gamma$ such that
the potential $V_+(z)$ is a double-well with well-separated wells (the form of this
potential is essentially determined by the atomic interaction parameter), the
oscillator approximation is still valid locally (i.e. about the two minima of the
potential $V_+(z)$, see Fig. \ref{FG1}) and the ground state can be now
approximated by a combination of the local eigenfunctions of the form (\ref{EQ28})
due to the quantum tunneling between the wells of  $V_+(z)$.

The case $\vare=0$  is the most interesting. The two local minima, solutions of Eq.
(\ref{EQ17}), read $z^{(\pm)}_s = \pm\sqrt{1-\gamma^{-2}}$. The validity of the
local approximation by oscillator eigenfunctions in the two wells, i.e. the
condition of well-separated wells of the potential $V_+(z)$ for $\gamma<-1$, is
defined by how much the width of the local oscillator eigenfunctions is smaller
than the distance between the two wells. We have $z_R - z_L =
2\sqrt{1-\gamma^{-2}}$ and
 using Eq. (\ref{EQ28}) $\langle (z -z_L)^2\rangle =  4\langle (x -x_L)^2\rangle=
N^{-1}|\gamma|^{-1}(\gamma^2-1)^{-1/2}$. Therefore, the applicability condition
$\langle (z -z_L)^2\rangle \ll (z_R-z_L)^2/4$ reads
\eqb
|\gamma|(|\gamma|-|\gamma|^{-1})^{3}\gg \frac{1}{N^2},
\label{EQ33}\eqe
where on the l.h.s. we have a monotonically growing function of $|\gamma|$. The
condition is not too restrictive and already for $|\gamma|$ slightly above the
critical value $|\gamma|=1$ it is satisfied, e.g. setting $|\gamma| = 1 + \delta$
with $\delta \ll1$ in Eq. (\ref{EQ33}) we obtain  $\delta \gg N^{-2/3}$ which for
$N=1000$ gives $\delta \gg 0.01$. Hence already $|\gamma| = 1+\delta \ge 1.1$
satisfies the condition (\ref{EQ33}) for $N=1000$.

To obtain the ground state we need to evaluate the tunneling rate, i.e. the matrix
element $\langle\Psi_L|\hat{H}|\Psi_R\rangle$, where for the localized
eigenfunctions in the left and right well of the double well $V_+(z)$ one can use
the local approximation (\ref{EQ28}) and $\hat{H}$ can be taken from Eq.
(\ref{EQ15}).  We have for the localized states in the left and right wells (here
$s = L,R$)
\eqb
\Psi_{s}(z) = \frac{1}{\pi^{1/4}\sqrt{\sigma}}e^{-\frac{(z-z_s)^2}{2\sigma^2}},
\label{EQ34}\eqe
where $\sigma^2 = 4\sigma_+^2 = \frac{2}{N}|\gamma|^{-1}(\gamma^2-1)^{-1/2}$. Note
that there is an exponentially small overlap between $\Psi_L$ and $\Psi_R$:
\begin{eqnarray}
\langle\Psi_L|\Psi_R\rangle &=& \frac{1}{\sqrt{\pi}\sigma}\int\limits_{-1}^1\rd
z\,\exp\{-\frac{(z-z_L)^2+(z-z_R)^2}{2\sigma^2}\}\nonumber\\
& = &\exp\left\{-\frac{N}{2}|\gamma|^{-1}(\gamma^2-1)^{3/2}\right\},
\label{EQ35}\end{eqnarray}
where we have used that $(z-z_L)^2+(z-z_R)^2 = 2(z^2+z_L^2)$ and the normalization
of the function $\Psi(z) = \pi^{-1/4}\sigma^{-1/2}\exp\{-\frac{z^2}{2\sigma^2}\}$.
Using the integration by parts we get
\begin{eqnarray*}
\langle\Psi_L|\hat{H}|\Psi_R\rangle &=&
\frac{2}{\sigma^4N^2}\langle\Psi_L|(z-z_L)(z-z_R)\sqrt{1-z^2}|\Psi_R\rangle
\\
&&+ \langle\Psi_L|\gamma z^2/2-\sqrt{1-z^2}|\Psi_R\rangle.
\end{eqnarray*}
To estimate the products we use  $(z-z_L)(z-z_R) = z^2 -1 +\gamma^{-2}$ and set
$z=0$ where the overlap of the wave-functions $\Psi_{L,R}(z)$ is maximal, with the
effect \footnote{We have checked numerically that discarding $z$-terms in the
integral for $\kappa$ gives sufficiently accurate results.}
\eqb
\langle\Psi_L|\hat{H}|\Psi_R\rangle \approx -\left[1 +
\frac{(\gamma^2-1)^2}{2}\right]\langle\Psi_L|\Psi_R\rangle.
\label{EQ36}\eqe
It is natural to define  the ``tunneling coefficient'' $\kappa$ for the effective
quantum particle as follows (cf. with Eq. (\ref{EQ2}))
\begin{eqnarray}\label{EQ37}
\kappa &=& - \langle\Psi_L|\hat{H}|\Psi_R\rangle \\
&\approx& \left[1 +
\frac{(\gamma^2-1)^2}{2}\right]\exp\left\{-\frac{N}{2|\gamma|}(\gamma^2-1)^{3/2}\right\}\nonumber.
\end{eqnarray}

\begin{figure}[htb]
\begin{center}
\psfig{file=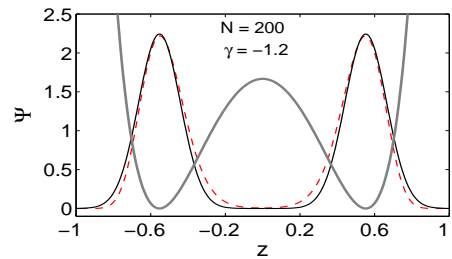,height=0.275\textwidth,width=0.35\textwidth} \caption{(Color
online) The ground state of attractive  BEC in the supercritical case: theoretical
(given by Eqs. (\ref{EQ34})  and (\ref{EQ39}), solid line) vs. numerical
diagonalization of the Hamiltonian (dashed line). The thick grey line shows
schematically  the effective potential $V_+(z)$. Here $N = 200$, $\gamma = -1.2$
and $\vare =0$. }
\label{FGCat}
\end{center}
\end{figure}

The ground and first excited states in the supercritical double-well potential
$V_+(z)$ for $\gamma<0$ read
\eqb
|G\rangle = \frac{1}{\sqrt{2}}(|\Psi_L\rangle + |\Psi_R\rangle),\; |E\rangle =
\frac{1}{\sqrt{2}}(|\Psi_L\rangle - |\Psi_R\rangle),
\label{EQ39}\eqe
where $\Psi_{L,R}(z)$ are given by Eq. (\ref{EQ34}). The theoretical ground state
is a very good approximation of the  numerical diagonalization of the quantum
Hamiltonian (\ref{EQ1}), see Fig. \ref{FGCat}.

\begin{figure}[htb]
\begin{center}
\psfig{file=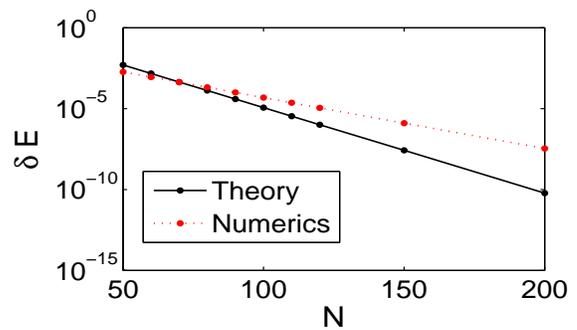,height=0.25\textwidth,width=0.425\textwidth} \caption{(Color
online) The energy splitting due to tunneling in the effective potential well vs.
number of atoms $N$: theoretical (solid line) and numerical  (dotted line). Here
$\gamma = -1.2$ and $\vare =0$. }
\label{FGdE}
\end{center}
\end{figure}

The degenerate level energy splitting, given by $\delta E = 2\kappa$, is a
qualitatively good result, see Fig. \ref{FGdE}. Note that the $N$-dependence of the
energy duplet splitting is manifestly exponential (however, the local approximation
used for the wave functions cannot capture the right coefficient in the exponent).
The energy splitting can be also estimated by using the perturbation theory, but
only for small values of $N$ (since the small parameter is equivalent to $N/\gamma$
see, for details, Ref. \cite{BES}).

The atom number fluctuations in the ground state (\ref{EQ39}) read
\eqb
\langle\Delta n_1^2\rangle = \frac{N^2}{4}\langle G|z^2|G\rangle \approx
\frac{N^2}{4}(1-\gamma^{-2}).
\label{EQ41}\eqe

For large $|\gamma|$ above the critical value the ground state of an attractive BEC
in the symmetric double-well trap is the Schr\"odinger cat-state (see also Refs.
\cite{BECcats1,BECcats2}). Precisely, for $|\gamma|\gg\sqrt{N}$ we have
\eqb
|G\rangle = \frac{1}{\sqrt{2}}(|N,0\rangle + |0,N\rangle).
\label{EQ40}\eqe
Indeed, in the supercritical case  $\Psi_{L,R}(z)$,  given by Eq. (\ref{EQ34}), has
the width in the Fock space given by $N\sqrt{\langle(z-z_s)^2\rangle}/2 =
\frac{\sqrt{N}}{2}|\gamma|^{-1/2}(\gamma^2-1)^{-1/4}\approx \sqrt{N}/(2|\gamma|)\ll
1$ for  $|\gamma|\gg\sqrt{N}$. Thus each state $|\Psi_{L,R}\rangle$ is a Fock
state.

\section{MQST and  the negative mass quantum particle }
\label{secVI}

The equivalence between the  repulsive and  attractive BEC cases, given by Eq.
(\ref{RepAttr}), means, for instance, that the repulsive BEC also contains the
Schr\"odinger cat-state (see also Ref. \cite{GE}), which is  an excited stationary
state.

The double-well potential $V=V_-(z)$ for the classical phase $\phi=\pi$  is
responsible for the  MQST states  of a repulsive BEC in a symmetric double-well,
predicted in Ref. \cite{MFDW} and observed experimentally  in Ref. \cite{TunTrap}.
There are two types of the MQST: the phase-locked ($\phi \approx \pi$) states  and
the running phase states ($\phi \propto t$). Consider the symmetric double-well
trap ($\vare=0$). Following the arguments of Ref. \cite{MFDW} in the classical
case, i.e. using $\mathcal{H} = E$ and Eq. (\ref{EQ19}) one arrives at the equation
\eqb
\frac{1}{4}\left(\frac{\rd z}{\rd \tau}\right)^2 =  1-z^2
-\left(\frac{\gamma}{2}z^2 - E\right)^2,
\label{EQ42}\eqe
which results in inaccessible  $z=0$ region for $|E|>1$. Since for  $N\gg1$ the
energy satisfies $V_+(z)\le E \le V_-(z)$, this is possible only when the potential
$V_-(z) = \frac{\gamma}{2}z^2+ \sqrt{1-z^2}$ has a local minimum at $z=0$ (i.e. it
is an inverted double-well) and the energy line crosses it. Hence, the MQST is
possible only for $\gamma>1$. The mean-field condition for the MQST reads $\gamma
>\gamma_{\mathrm{c}}$ where \cite{MFDW}
\eqb
\gamma_{\mathrm{c}} = 2\frac{1 + \sqrt{1-z^2(0)}\cos[\phi(0)]}{z^2(0)}.
\label{MQSTcond}\eqe
One can easily verify that the mean-field critical value always satisfies
$\gamma_{\mathrm{c}}>1$, which is easily seen by  rewriting  Eq. (\ref{MQSTcond})
as $\gamma z(0)^2/2 = 1 +\sqrt{1-z(0)^2}\cos{\phi(0)}$ and noticing that if
$\gamma\le 1$ the functions on the l.h.s. and on the r.h.s. have no intersections
for $0<|z(0)|<1$. Note that for $\cos\phi(0)\ge0$ we get $\gamma_\mathrm{c}\ge2$.
For $\gamma>2$ the MQST condition reads $|z(0)|>z_c$ where $z_c$ is the solution of
Eq. (\ref{MQSTcond}). On the other hand, for $\cos\phi(0)<0$ and $1 + |\sin\phi(0)|
<\gamma<2$ there is an interval of the initial population imbalance for MQST:
$z_{1} < |z(0)|<z_{2}$, where $z_{1,2}$ solve Eq. (\ref{MQSTcond}).

\begin{figure}[htb]
\begin{center}
\psfig{file=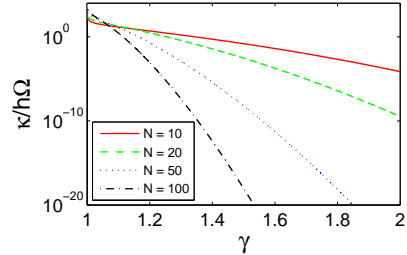,height=0.2\textwidth,width=0.35\textwidth} \caption{(Color
online) An illustration of the   quantum condition, i.e. $\kappa/h\Omega_-\ll 1$,
for the phase-locked MQST states for several values of the number of BEC atoms $N$.
}
\label{FG3}
\end{center}
\end{figure}

If repulsive BEC is prepared in one well of the double-well trap and the initial
phase $\phi(0)\approx\pi$ one can use the negative mass Schr\"odinger equation
(\ref{EQ15}) to explain the quantum dynamics. In this case the quantum potential in
the Fock space $V = V_-(z)$  has the  double-well form with the quasi-degenerate
energy levels, which  reflect the existence of  two classical fixed points
$z_s^{(\pm)}=\pm\sqrt{1-\gamma^{-2}}$. The degeneracy is estimated as twice the
tunneling coefficient of Eq. (\ref{EQ37}). Therefore, the observation of the
mean-field MQST with phase $\langle \phi \rangle \approx \pi$ is subject to the
quantum condition that the oscillation time of the effective quantum particle in
one of the  wells is much less than the tunneling time between the wells of the
double-well potential $V_-(z)$, i.e. in the energy terms $2\kappa\ll h \Omega_-
=2h\sqrt{a_-b_-}$ or
\eqb
\left[1+\frac{(\gamma^2-1)^2}{2}\right]\exp\left\{-\frac{N}{2\gamma}(\gamma^2-1)^{3/2}\right\}\ll
\frac{\sqrt{\gamma^2-1}}{N},
\label{EQ43}\eqe
which, as is demonstrated in Fig. \ref{FG3}, is  satisfied for all $\gamma$ just
above the critical value,  even for small number of BEC atoms. This fact makes
possible the experimental observation of the phase-locked mean-field MQST.  A
similar conclusion was made before using different approach \cite{ST,RSK}. We note
also that the $\pi$-phase MQST was analyzed in Ref. \cite{ST} by considering the
numerical eigenvalue spectrum  (see also Ref. \cite{GE}) and  was related to the
appearance of the quasi-degenerate energy doublets.

\begin{figure}[htb]
\begin{center}
\psfig{file=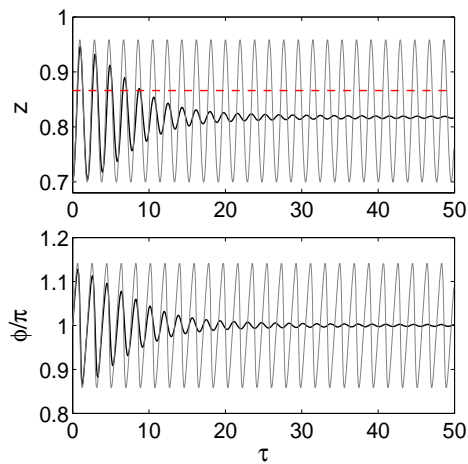,height=0.35\textwidth,width=0.35\textwidth} \caption{(Color
online) The phase-locked MQST state dynamics: quantum averages (black lines) vs
classical dynamics (grey lines). The dashed line in the upper panel gives the
classical stationary point $z_s = 0.866$. Here $N = 1000$, $\gamma = 2$ (whereas
$\gamma_{\textrm{c}}= 1.23$) and $\vare = 0$. The initial Gaussian distribution of
width $\Delta z = 0.12$ centered at the point $z_0=0.78$ and with phase $\phi =
\pi$ was used. }
\label{FG4}
\end{center}
\end{figure}

\begin{figure}[htb]
\begin{center}
\psfig{file=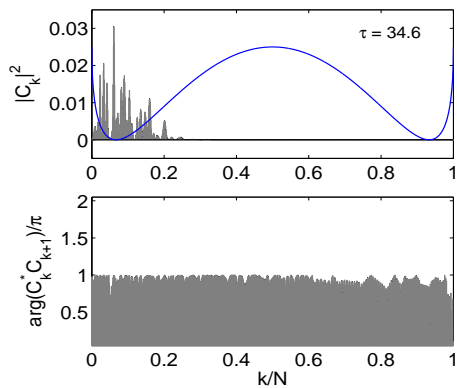,height=0.35\textwidth,width=0.35\textwidth} \caption{(Color
online) The atomic number distribution (a) (represented by vertical bars) of the
phase-locked MQST state and the distributed quantum phase (b) (given by bars).  The
line in panel (a) gives a schematic portrait of the inverse quantum potential
$-V_-(1-2k/N)$. }
\label{FG5}
\end{center}
\end{figure}

Numerical simulations show that the quantum average values defined as \footnote{The
usual definition of the phase is via the average $\langle a^\dag_1a_2\rangle$. Our
definition is slightly different and prompted by the quantum-classical
correspondence established in section \ref{secII}: $\hat{p}\to \phi$. The two
definitions agree very well for large $N$. Further, we use $e^{i\hat{p}}$  to
define the dispersion, since  is defined for any $N$, while $\hat{p}$ is
ill-defined for small $N$. }
\eqb
\!\!\!\langle z\rangle = 1 - \frac{2}{N}\sum_{k=0}^N k|C_k|^2, \; \langle
\phi\rangle = \textrm{arg}\langle e^{i\hat{p}}\rangle =
\textrm{arg}\sum_{k=0}^{N-1}C^*_kC_{k+1},
\label{EQ44}\eqe
initially follow  the mean-field dynamics, see Fig. \ref{FG4}. The atomic
distribution in the Fock space remains localized in one of the wells of the inverse
quantum potential $-V_-(z)$, Fig. \ref{FG5}(a). The distributed quantum phase
$\phi_k$, defined as $\phi_k = \textrm{arg}(C^*_kC_{k+1})$ remains very close to
$\pi$, see Fig. \ref{FG5}(b). For long times the classical oscillations are subject
top collapses and revivals, see Ref. \cite{MCWW}. The numerical method of
propagating  the Schr\"odinger equation  is adopted  from Ref. \cite{NMeth} (see
also Ref. \cite{SK3}).

To get a quantitative estimate of the validity  the classical dynamics, one can use
the standard deviations of the quantum variables, defined as \mbox{$\Delta z =
2\Delta x = 2|| x - \langle x\rangle||=\left[\langle(x - \langle
x\rangle)^2\rangle\right]^{\frac12}$}  and
\begin{eqnarray}
\Delta e^{i\hat{p}} &=& ||e^{i\hat{p}} - \langle e^{i\hat{p}}\rangle||
\nonumber\\
&=&\left[\langle\left(e^{-i\hat{p}} - \langle e^{-i\hat{p}}\rangle
\right)\left(e^{i\hat{p}}
- \langle e^{i\hat{p}}\rangle\right)\rangle\right]^{\frac12}\nonumber\\
&=&\left[1 - |\langle e^{i\hat{p}}\rangle|^2\right]^{\frac12}
\label{EQ45}\end{eqnarray}
(we assume that $|\langle N,0|\Psi\rangle|=0$, in the case of $|\langle
N,0|\Psi\rangle|\ne0$ and  $|\langle 0,N|\Psi\rangle|=0$ one can use the operator
$e^{-i\hat{p}}$ instead with similar result). From Eq. (\ref{EQ45}) one concludes
that for the average phase defined in Eq. (\ref{EQ44}) we have $\langle
e^{i\hat{p}}\rangle = \alpha e^{i\langle\phi\rangle}$ with some $\alpha\le 1$. The
quantum dispersions satisfy the  uncertainty relation
\eqb
\frac{\Delta e^{i\hat{p}}}{|\langle e^{i\hat{p}}\rangle|}\Delta z \ge h.
\label{INEQ}\eqe
For the simulations presented in figure \ref{FG4}, the standard deviation $\Delta
z$ decreases from $0.122$ settling to  $0.096$ while  $\Delta e^{i\hat{p}}$ grows
from $0.0479$ to $0.268$ (the average $|\langle e^{i\hat{p}}\rangle|$ decreases
from $0.999$ to $0.963$). In this case the l.h.s. in the inequality (\ref{INEQ})
grows in the result of evolution by less than one order of magnitude as compared to
the initial value  on the order of the r.h.s..

\begin{figure}[htb]
\begin{center}
\psfig{file=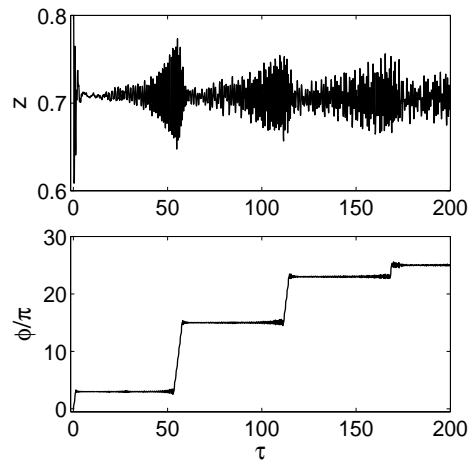,height=0.35\textwidth,width=0.35\textwidth} \caption{The
quantum collapses and revivals of the running phase MQST state. We use $N=200$,
$\gamma = 5$, $\vare =0$. The initial condition is a Gaussian with the average $z_0
= 0.91$ (the $y$-axis is cut below the upper limit of $z$ for better visibility of
the dynamics at large times),  $\phi(0)=0$  (thus $\gamma_\mathrm{c} = 3.35$) and
the Gaussian  width $\sigma=0.1N$.}
\label{FG6}
\end{center}
\end{figure}

\begin{figure}[htb]
\begin{center}
\psfig{file=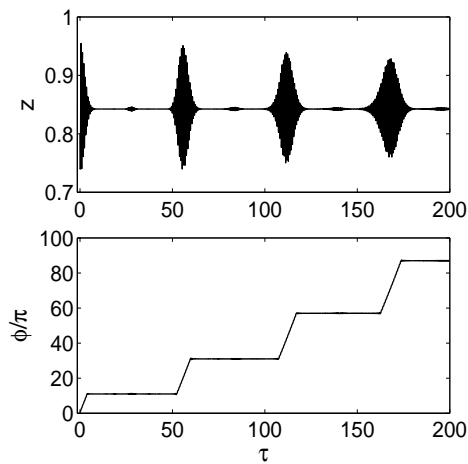,height=0.35\textwidth,width=0.35\textwidth} \caption{The same
as in Fig \ref{FG6} ($N=200$, $\gamma = 5$, $\vare =0$) for the Gaussian initial
condition with the average $z_0 = 0.8$ and the initial phase $\phi(0) = 0.75\pi$
(thus $\gamma_\mathrm{c} = 1.8$) and width $\sigma=0.05N$.}
\label{FG6A}
\end{center}
\end{figure}

It was the running phase MQST which was observed in the experiment of Ref.
\cite{TunTrap}. The classical dynamics of the running phase MQST is well understood
\cite{MFDW}. We have found that the quantum correction to the classical running
phase is in the form of the quantum collapses and revivals, see Figs. \ref{FG6} and
\ref{FG6A}. Though we show the numerical simulations for small number of atoms
$N=200$, the time of occurrence of the first collapse of the running phase does not
seem to depend on the number of atoms but on the initial conditions, for instance
the phase (the subsequent quantum revivals and collapses do depend on $N$: we have
found no revivals for $N=1000$ up to $\tau = 200$). This effect can be observed in
an experiment if the dynamics is followed for longer times.

\begin{figure}[htb]
\begin{center}
\psfig{file=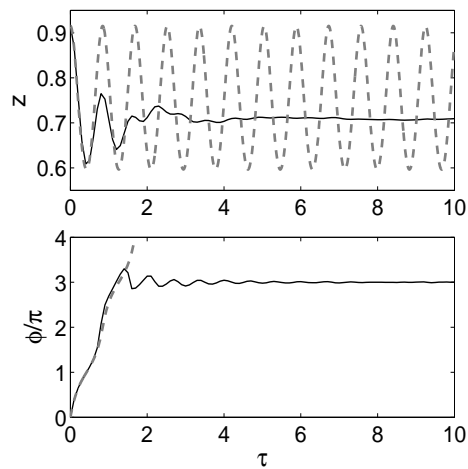,height=0.35\textwidth,width=0.35\textwidth} \caption{The
classical dynamics (thick dashed lines) and the corresponding quantum averages
(thin solid lines) at the first occurrence of quantum collapse of the running phase
MQST state of Fig. \ref{FG6}.  }
\label{FG7}
\end{center}
\end{figure}

\begin{figure}[htb]
\begin{center}
\psfig{file=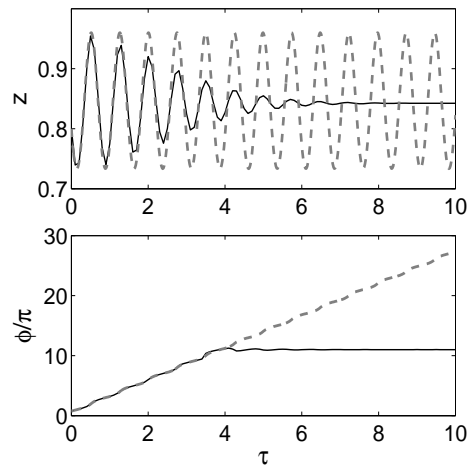,height=0.35\textwidth,width=0.35\textwidth} \caption{The same
as in Fig. \ref{FG7} but for $\phi(0)=0.75\pi$, i.e. corresponding to Fig.
\ref{FG6A}.  }
\label{FG7A}
\end{center}
\end{figure}

For $\phi(0)=0$ the first collapse of  the running phase occurs about the value
$\phi(\tau) = 3\pi$, which is close to final value of the phase in the experiment
of Ref.~\cite{TunTrap} on the MQST (this depends on the initial conditions: for
$\phi(0)=0.75\pi$ the first collapse occurs at $\phi(\tau) \approx 11\pi$, see Fig.
\ref{FG7A}). The growth of the quantum average $\langle \phi\rangle$ is interrupted
by plateaus of constant phase, while the classical phase follows the linear growth,
see Figs. \ref{FG7} and \ref{FG7A}. In the case of $\phi(0)=0.75\pi$, Figs.
\ref{FG6A} and \ref{FG7A}, the initial state is more classical, i.e. the
corresponding uncertainty relation (\ref{INEQ}) is approximately equality at $t=0$
(whereas in the case of $\phi=0$, initially, the l.h.s of Eq. (\ref{INEQ}) is
larger than the r.h.s. by an order of magnitude).

\begin{figure}[htb]
\begin{center}
\psfig{file=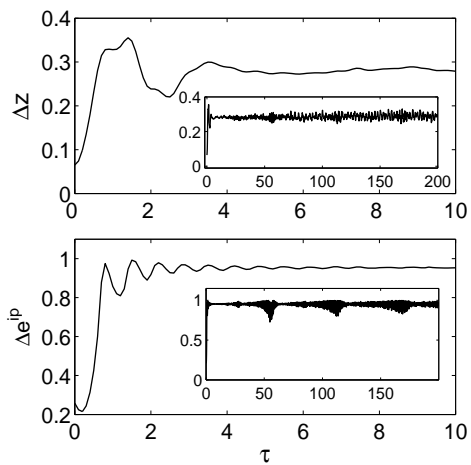,height=0.35\textwidth,width=0.35\textwidth} \caption{The
dispersions of the quantum distributions of $z$ and $e^{i\hat{p}}$  corresponding
to Figs. \ref{FG6} and  \ref{FG7}. The insets show the full  picture of  time
dependence of the dispersions.}
\label{FG8}
\end{center}
\end{figure}

\begin{figure}[htb]
\begin{center}
\psfig{file=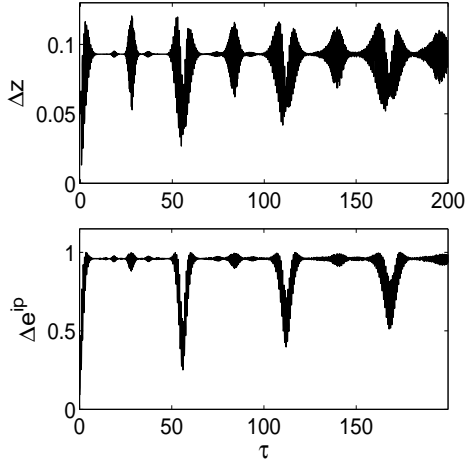,height=0.35\textwidth,width=0.35\textwidth} \caption{The
dispersions of the quantum distributions of $z$ and $e^{i\hat{p}}$ corresponding to
Figs. \ref{FG6A} and \ref{FG7A}. }
\label{FG8A}
\end{center}
\end{figure}

Our principal result is that the first quantum collapse is associated with an
exponential  growth of quantum fluctuations of the phase distribution, see Figs.
\ref{FG8} and \ref{FG8A}, which reach a maximal value at the first occurrence of
the quantum collapse. In this case, the l.h.s. of the inequality (\ref{INEQ}) grows
by more than two orders of magnitude reaching the value of order one at the first
collapse (the average $|\langle e^{i\hat{p}}\rangle|$ decreases from $0.97$ to
$0.03$). The growing fluctuations are also seen in the experimental results on the
running phase MQST presented in Ref. \cite{TunTrap}.

Finally, we note that  the running phase  MQST  can be described by the WKB
approach by setting in Eq. (\ref{EQ8})  $\phi = \phi(\tau)$ (i.e. the classical
solution) and solving the time-dependent Schr\"odinger equation analogous of Eq.
(\ref{EQ12}). In this approach, it is the running classical phase what makes the
potential (\ref{Vh}), seen by the effective quantum particle, asymmetric in $x$,
since now we have $\mathcal{V}_{MQST} = \mathcal{V}_0(x,\phi) + \dot{\phi}x$. This
gives an elementary  explanation why BEC is trapped in one well of the double-well
trap.

\section{Conclusion}
\label{secVII}

We have proposed an analytical approach for description of quantum phenomena in the
system of large number of interacting identical bosons occupying only few modes
(two in the present study). The method links the many-boson system  with the
dynamics of a single quantum particle in a potential, where the normalized
occupation  numbers of the different modes in the Fock space serve as the particle
co-ordinates. We have used as the example the well-known two-mode model,
describing, for instance, BEC tunneling in a double-well trap, i.e. the boson
Josephson effect. The method allowed to account for the  mean-field stationary
points and their stability, trace analytically the transformation of the ground
state of the system of identical bosons in both attractive and repulsive case,
derive the quantum fluctuations of the number of atoms in the ground state, relate
the appearance of the macroscopic quantum self-trapping phenomenon to the
double-well shape of the potential for the effective quantum particle and give a
quantum explanation of the phase-locked and running phase self-trapped states of
BEC in the double-well trap. We also predict a new phenomenon  -- quantum collapses
and revivals of classical running phase of the macroscopic quantum self-trapped
state.

Our method awaits other important applications, where few-mode boson models
naturally appear, including the theory of molecular-atomic coherence in BEC
\cite{AMTheory}, the quantum model of nonlinear intraband tunneling of BEC in
optical lattices \cite{SK2,SK3} and many others.

\acknowledgments   V.S.S. thanks  the CAPES of Brazil for partial financial
support.

\begin{appendix}

\section{The full two-mode boson-Josephson  model }
\label{appH}

One can show that the full two-mode boson model describing BEC in a double-well
trap can be cast in the (dimensional) form
\begin{eqnarray}
H &=& \delta V a_1^\dag a_1- [J_\mathrm{lin} +J_\mathrm{nonl}(N-1)](a_1^\dag a_2  + a_2^\dag a_1) \nonumber\\
& +&\frac{\rho_1-\rho_2}{2}\left[(a^\dag_1a_1)^2
+(a^\dag_2a_2)^2\right]+\frac{\rho_2}{2}\left(a_1^\dag a_2  + a_2^\dag
a_1\right)^2,\nonumber\\
\label{ApH1}\end{eqnarray}
where some scalar $N$-dependent term  has been  discarded. The coefficients are
given as $J_\mathrm{lin} = \delta E/2$,
\begin{eqnarray}
&&J_\mathrm{nonl} = -g\int\rd^3\, \bx\varphi_{\alpha}(\bx)\varphi_{\beta}^3(\bx)
,\nonumber\\
&&\rho_1= g\int\rd^3 \bx\,\varphi_{\alpha}^4(\bx),\nonumber\\
 &&\rho_2 = g\int\rd^3\, \bx\varphi_\alpha^2(\bx)\varphi_\beta^2(\bx)
\label{ApH2}\end{eqnarray}
(the subscripts $\alpha$ and $\beta$ are permutation of the list $\{L,R\}$ and the
functions give the localized states in the  left or right well defined by the
appropriate linear combinations of the ground state and the first exited state, see
section \ref{secII}). The derivation is similar to that of Refs. \cite{AB,GO} and
is omitted. However, with the help of numerical evaluation, one can verify that for
all double-well traps with two lower degenerate levels $E_1$ and $E_2$ satisfying
the inequality $\delta E = E_2-E_1 \ll E_3 - E_2$ the coefficients satisfy
\eqb
\rho_1 \gg \rho_2\gg |J_\mathrm{nonl}|.
\label{ApH3}\eqe
The coefficients $J_\mathrm{lin}$ and $J_\mathrm{nonl}(N-1)$, however, can be of
the same order. Discarding the small terms and dividing the Hamiltonian by the
quantity $[J_\mathrm{lin}+J_\mathrm{nonl}(N-1)]N$ one gets the reduced model given
by Eq. (\ref{EQ1}) with a  different definition of the parameters.

\section{Proof that the local minimum of $-V_-(z)$ at the negative $z$ is  the absolute
minimum}
\label{appA}

Denote $z_-$ and $z_+$ the negative and positive local minima of $-V_-(z)$ for
$\gamma>\gamma_{\mathrm{cr}}$. Using \mbox{$\vare z = z^2(\gamma-(1-z^2)^{-1/2})$}
from Eq. (\ref{EQ17}) into the expression for the quantum potential (\ref{EQ16}) we
get
\eqb
V_-(z_\pm)=  \frac{z_\pm^2}{2}\left( \frac{2}{\sqrt{1-z_\pm^2}}-\gamma\right)
+\sqrt{1-z_\pm^2}.
\label{A1}\eqe
Using Eq. (\ref{EQ17}) again to  obtain $(1-z_-^2)^{-1/2}\ge\gamma$ and
$(1-z_+^2)^{-1/2}\le\gamma$, we arrive at the estimates
\eqb
V_-(z_+) \le \frac{1-z_-^2/2}{\sqrt{1-z_-^2}},\quad  V_-(z_-) \ge
\frac{1-z_-^2/2}{\sqrt{1-z_-^2}},
\label{A2}\eqe
where the equality sign is for $\vare =0$. Now, since the function $f(z) \equiv
(1-z^2/2)/\sqrt{1-z^2}$ is monotonously growing for $0<z<1$, one needs to compare
just the values $|z_-|$ and $z_+$. But it is evident from Fig. \ref{FG1}(b) that
$z_+\le|z_-|$ (the equality for $\vare = 0$) which leads to $V_-(z_-)\ge V_-(z_+)$.
Indeed, $z_\pm$ are the intersections of the curve
$\mathrm{lhs}(z)\equiv\gamma-(1-z^2)^{-1/2}$ with the two branches of $\vare/(2z)$,
while its intersections $z^{(\pm)}_0$ with  the $z$-axis  are equal in the absolute
value. But, since the derivative of $\mathrm{lhs}(z)$ has the  sign opposite that
of $z$, we get $z_-<z_0^{(-)}$ and $z_+<z_0^{(+)}$ what completes the proof.

\section{The details of the analysis of the ground state of BEC}
\label{appB}

To find the ground state of large BEC ($N\gg1$) one needs to compare just the
zero-point energies at the extremal points (which is the classical energy of the
stationary point, Eqs. (\ref{EQ20}) and (\ref{EQ21})). Using Eq. (\ref{EQ17}) we
get
\eqb
V_\pm(z_{\pm}) = -\frac{\gamma z_{\pm}^2}{2} \mp\frac{1}{\sqrt{1-z_{\pm}^2}},
\label{EQ24}\eqe
where $z_{\pm}=z_s^{(\pm)}$ is the corresponding  extremal point of $V_\pm(z)$.

Consider the symmetric case $\vare = 0$ and a repulsive BEC $\gamma>0$. The
extremal points are  $z_+ = 0$ and $z_-=0$ for $\gamma<\gamma_\mathrm{cr}=1$, while
$z^2_- = 1-\gamma^{-2}$ otherwise. Using them  into Eq. (\ref{EQ24}) we get
\eqb
V^{(\mathrm{rep})}_+(z_+) = -1,\quad  V^{(\mathrm{rep})}_-(z_-) = \left\{  \begin{array}{c} 1,\;\gamma\le1\\
\frac{1+\gamma^2}{2\gamma}, \; \gamma>1 \end{array} \right.
\label{EQ25}\eqe
Thus the ground state  of a repulsive BEC in the double-well trap is given by the
equal distribution of atoms between the wells with the zero phase difference.

On the other hand, for an attractive BEC in a symmetric double-well trap, using Eq.
(\ref{EQ24}) we obtain $z_-=0$ and $z_+ = 0$ for $|\gamma|\le1$, while $z^2_+ =
1-\gamma^{-2}$ otherwise. Eq. (\ref{EQ24}) then gives
\eqb
V^{(\mathrm{attr})}_+(z_+) = \left\{  \begin{array}{c} -1,\;|\gamma|\le1\\
\frac{1+\gamma^2}{2\gamma}, \; |\gamma|>1 \end{array} \right.,\quad
V^{(\mathrm{attr})}_-(z_-) = 1,
\label{EQ26}\eqe
where now $\gamma<0$. Thus the ground state of an attractive BEC also has equal
distribution of atoms between the wells of the double-well trap (with zero phase
difference)  for $|\gamma|<1$. On the other hand, the ground state of an attractive
BEC for $|\gamma|>1$, when there are  two local bound states (in the classical case
$z_s>0$ and $z_s<0$) corresponding to unequal distributions, is given by the ground
state $\psi(z)$ of the effective particle in the double-well trap $V_+(z)$, see Eq.
(\ref{EQ15}). In this case, the classical stationary points feature the spontaneous
symmetry breaking.

Finally, since for arbitrary $\vare>0$ the only change is in the position of
$z_\pm$ and Eq. (\ref{EQ24}) is derived for arbitrary $\vare$, the ground state of
BEC for $\vare\ne0$  is given by the ground state of Eq. (\ref{EQ14}) with positive
mass. Indeed, we estimate from Eq. (\ref{EQ24}) using (\ref{EQ17}):
\begin{eqnarray}
&& V_+(z_+) < -\frac{z_+^2}{2}\left(\gamma + \frac{1}{\sqrt{1-z_+^2}}\right) =
-\frac{\vare z_+}{4},\nonumber\\
&& V_-(z_-) > \frac{z_-^2}{2}\left(\gamma  + \frac{1}{\sqrt{1-z_-^2}}\right) =
-\frac{\vare z_-}{4}.
\label{EQ27}\end{eqnarray}
We arrive at the needed inequality $V_+(z_+)<V_-(z_-)$ for all $\vare$, since for a
repulsive BEC the absolute minimum of $-V_-(z)$ is negative $z_-<0$ (see Fig.
\ref{FG1}(a)) while $z_+ >0$ Eq. (\ref{EQ17}). On the other hand, by Eqs.
(\ref{EQ17}) and (\ref{RepAttr}), for an attractive BEC $z_+$ is the absolute
minimum of the double well $V_+(z)$ (see Fig \ref{FG1}(a)) and is positive, while
$z_-$ is negative.
\medskip

\section{The relative atom number fluctuations in the positive mass case}
\label{appC}

We have $\langle (x-x_s)^2\rangle = \sigma^2_+/2$, where
\eqb
\sigma^2_+ = \frac{1}{2N}\left[\frac{(1-z_s)^{1/2}}{\gamma +
(1-z_s^2)^{-3/2}}\right]^{1/2}.
\eqe
From Eq. (\ref{EQ17}) for positive mass case we get $(1-z_s^2)^{1/2} =
\frac{\vare}{2z_s}-\gamma$. Now using  Eq. (\ref{zs}) we obtain:
\begin{eqnarray*}
&& \frac{\vare}{2z_s}-\gamma = 1 + \frac{\vare^2}{8(1+\gamma)^2} +O(\vare^4),
\\
&& \gamma + \left(\frac{\vare}{2z_s}-\gamma\right)^3 = (1+\gamma)\left[1 +
\frac{3\vare^2}{8(1+\gamma)^3} +O(\vare^4)\right].
\end{eqnarray*}
Therefore
\eqb
\sigma^2_+ = \frac{1}{2N}(1+\gamma)^{-1/2}\left[ 1 -
\frac{\vare^2(4+\gamma)}{16(1+\gamma)^3} +O(\vare^4)\right].
\eqe

\end{appendix}

\end{document}